\documentclass{article}

\usepackage{PRIMEarxiv}

\usepackage[utf8]{inputenc} % allow utf-8 input
\usepackage[T1]{fontenc}    % use 8-bit T1 fonts
\usepackage{hyperref}       % hyperlinks
\usepackage{url}            % simple URL typesetting
\usepackage{booktabs}       % professional-quality tables
\usepackage{amsfonts}       % blackboard math symbols
\usepackage{nicefrac}       % compact symbols for 1/2, etc.
\usepackage{microtype}      % microtypography
\usepackage{lipsum}
\usepackage{fancyhdr}       % header
\usepackage{graphicx}       % graphics
\usepackage{multirow}
\graphicspath{{media/}}     % organize your images and other figures under media/ folder

\title{Automated WBRT Treatment Planning via Deep Learning Auto-Contouring and Customizable Landmark-Based Field Aperture Design
\thanks{\textit{\underline{Citation}}: 
\textbf{Preprint. Under review.}}%Yao Xiao, et al. Title. Pages.... DOI:000000/11111.}} 
}

\author{
   Yao Xiao \\
   Department of Radiation Physics \\
   the University of Texas MD Anderson Cancer Center \\
   Houston, TX, United States \\
   \texttt{yxiao4@mdanderson.org} \\
   \And
   Carlos Cardenas \\
   Department of Radiation Oncology \\
   University of Alabama-Birmingham \\
   Birmingham, AL, United States\\
   \AND
   Dong Joo Rhee, Tucker Netherton, Lifei Zhang, Callistus Nguyen, Raphael Douglas, Raymond Mumme \\
   Department of Radiation Physics \\
   the University of Texas MD Anderson Cancer Center \\
   Houston, TX, United States \\
   \And
   Stephen Skett, Tina Patel \\
   Department of Medical Physics \\
   Guy's \& St Thomas NHS Foundation Trust \\
   London, United Kingdom  \\
   \And
   Chris Trauernicht \\
   Division of Medical Physics \\ 
   Stellenbosch University and Tygerberg Academic Hospital \\ Stellenbosch, South Africa  \\
   \And
   Caroline Chung \\
   Department of Radiation Oncology \\
   the University of Texas MD Anderson Cancer Center \\ 
   Houston, TX, United States \\
   \And
   Hannah Simonds \\
   Division of Radiation Oncology \\ 
   Stellenbosch University \\
   Stellenbosch, South Africa \\
   \And
   Ajay Aggarwal \\
   Department of Medical Physics \\ 
   Guy’s \& St Thomas NHS Foundation Trust \\
   London, United Kingdom \\
   \And
   Laurence Court \\
   Department of Radiation Physics \\
   the University of Texas MD Anderson Cancer Center \\ 
   Houston, TX, United States \\
   \texttt{LECourt@mdanderson.org} \\
}

\begin{document}
\maketitle

\begin{abstract}
\textbf{Purpose}: To develop and evaluate an automated whole-brain radiotherapy treatment planning pipeline with a deep learning-based auto-contouring and customizable landmark-based field aperture design.

\textbf{Methods}: The pipeline consisted of the following steps: (1) Auto-contour normal structures on computed tomography (CT) scans and digitally reconstructed radiographs using deep learning techniques, (2) locate the landmark structures using the beam’s-eye-view, (3) generate field apertures based on eight different landmark rules addressing different clinical purposes and physician preferences. Two parallel approaches for generating field apertures were developed for quality control. The performance of the generated field shapes and dose distributions were compared with the original clinical plans. The clinical acceptability of the plans was assessed by five radiation oncologists from four hospitals. 

\textbf{Results}: The performance of the generated field apertures was evaluated by Hausdorff Distance (HD) and Mean Surface Distance (MSD) from 182 patients’ field apertures used in the clinic. The average HD and MSD for the generated field apertures were 16±7 mm and 7±3 mm for the first approach, respectively, and 17±7 mm and 7±3 mm, respectively, for the second approach. The differences regarding HD and MSD between the first and the second approaches were 1±2 mm and 1±3 mm, respectively. A clinical review of the field aperture design, conducted using 30 patients, achieved a 100\% acceptance rate for both the first and second approaches, and the plan review achieved a 100\% acceptance rate for the first approach and a 93\% acceptance rate for the second approach. The average acceptance rate for meeting lens dosimetric recommendations was 80\% (left lens) and 77\% (right lens) for the first approach, and 70\% (both left and right lenses) for the second approach, compared with 50\% (left lens) and 53\% (right lens) for the clinical plans. 

\textbf{Conclusion}: This study provided an automated pipeline with two field aperture generation approaches to automatically generate whole-brain radiotherapy treatment plans. Both quantitative and qualitative evaluations demonstrated that our novel pipeline was comparable with the original clinical plans.

\end{abstract}

% keywords can be removed

\section{Introduction}
Brain metastases are the most common intracranial malignancies in the adult population, contributing to 308,102 new cases and 251,329 deaths worldwide based on the reports in the global cancer statistics 2020.~\cite{sung2021global} More than 40\% of patients with cancer develop brain metastases.~\cite{sperduto2020survival} Whole-brain radiotherapy (WBRT) treatment is a well-established treatment for patients with brain metastases by radiologically controlling both visible tumors and invisible micro-metastases.~\cite{tsao2012radiotherapeutic}

Conventional WBRT treatment planning has many challenges. First, it is time-consuming to obtain input from physicians, physicists, and dosimetrists regarding contours and field shape setups, including multileaf collimator (MLC) blocking. After the completion of CT simulation, it can take several hours to a day to obtain the input from all members of the radiotherapy team needed to start treatment.~\cite{agazaryan2020timeliness} In addition, institutions vary in their clinical approaches to WBRT treatment planning. Furthermore, limited resources in low-to-middle-income countries can lead to delays, especially in regard to human resources.~\cite{shah2019cancer} For these reasons, automation may improve the efficiency of WBRT, and it may enable medical staff to refocus their efforts on developing more complex treatment plans. It is essential, however, for automated WBRT to be customizable for individual clinical practices.~\cite{kao2015tumor,jiang2019dosimetric}

With the development of Artificial Intelligence (AI) techniques, automation of the WBRT treatment planning process has greatly simplified clinical workflows and improved the quality of treatment planning.~\cite{lin2020automated} The WBRT planning process can be automated by using deep learning to predict field apertures.~\cite{han2021clinical} In Han et al.'s work, DeepLabV3+ architecture was trained to automatically define the beam apertures on lateral-opposed fields using digitally reconstructed radiographs (DRRs).~\cite{han2021clinical} However, the approach lacks flexibility; it cannot be configured to suit local clinical practices. By providing anatomical landmarks as the rules, a configurable solution can address the limitation of lacking flexibility. As radiation safety is critical to patients undergoing radiation therapy, quality assurance (QA) is a common way to detect potential errors or plan failures and alert people in advance. Thus, to ensure treatment planning quality, we developed a secondary approach that can equally be performed in generating radiation field apertures as the first approach.

\section{Methods}
We developed two parallel automated approaches for generating field apertures. The approaches rely on anatomic landmarks in the beam’s-eye-view but differ in how these landmarks are generated. The first approach generates landmarks by auto-contouring various structures on the 3D CT images and then projecting them into the beam’s-eye-view. The second approach auto-contours the same structures directly in the beam’s-eye-view using 2D DRRs. Two approaches were developed so that one could be used for automated planning, and one could be used to verify the first approach, similar to automated QA approaches described preciously.~\cite{rhee2019automatic,kisling2020automatic} In both approaches the beam’s-eye-view contours are used to guide aperture creation.

\subsection{Auto-Contouring}
Structures for defining the anatomic landmarks (the brain, eyes, lenses, and C1 and C2 vertebral bodies) were contoured by deep learning-based auto-contouring. For the first approach, 3D CT was used in auto-contouring, where the brain, eyes, and lenses were obtained using Rhee et al.'s Head and Neck model~\cite{rhee2019automatic}, and vertebral bodies (C1 and C2) were obtained by using models by Netherton et al~\cite{netherton2020evaluation}. The contours were then projected into a right-lateral beam’s-eye-view. For the second approach, beam’s-eye-view DRRs were generated based on the 3D CT image set (270-degree gantry angle), and deep learning models (FCN-8s~\cite{long2015fully}) were developed using 887, 194, and 182 patients’ scans as the training, validation, and testing sets, respectively for segmenting the structures using the beam’s-eye-view DRRs. 

\subsection{Customizable Landmark-Based Field Aperture Design}

\begin{figure}
  \centering
  \includegraphics[width=0.5\textwidth]{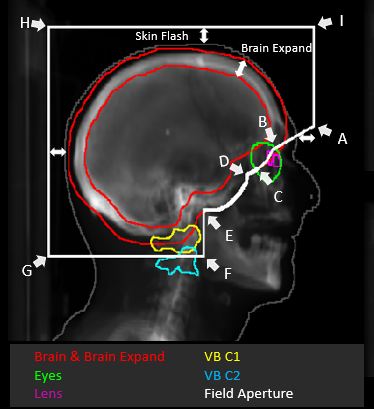}
  \caption{Field aperture design using different landmarks.}
  \label{fig:FieldDesign}
\end{figure}

The landmarks are the locations or points used to define the boundaries of the field apertures. In this work, landmarks were defined by the anatomic locations of tissues on the beam’s-eye-view. As shown in Figure ~\ref{fig:FieldDesign}, nine landmarks (A-I) were calculated to form the cranial (HI), caudal (FG), anterior (AI, EF), posterior, and anterior-caudal (AB, BC, CD, and DE) boundaries of the radiation field aperture.

\begin{figure}
  \centering
  \includegraphics[width=\textwidth]{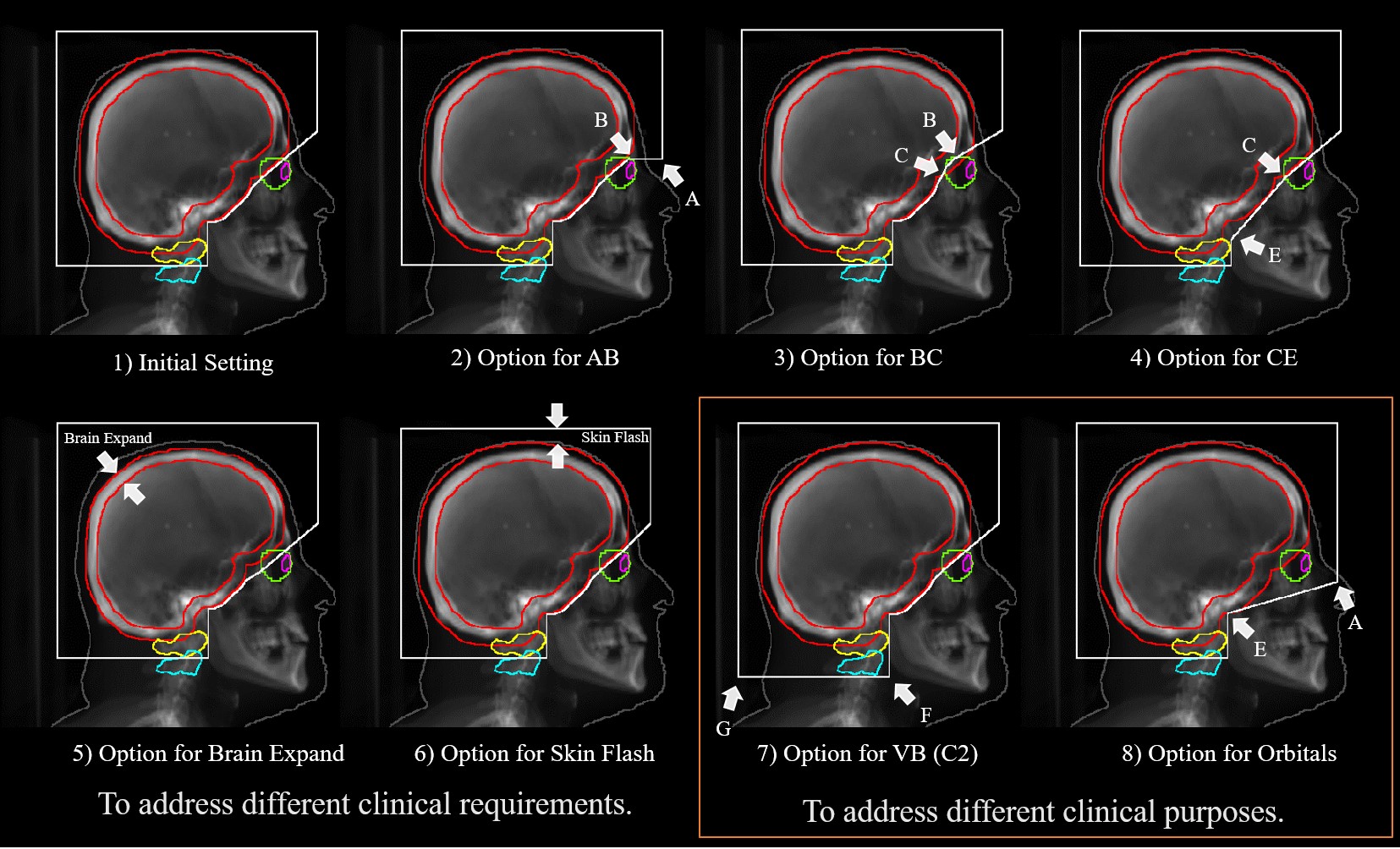}
  \caption{Generated field apertures based on different options. Options 1-6 are configured to address different clinical preferences and options 7 and 8 are configured to address different clinical purposes. 1). The initial setting. 2) The line shapes for AB (horizontal or slash/diagonal-like. 3). The option of adjusting the line BC at different distances between the eyes and the cribriform plates (close to the lenses or eyes at the cranial-posterior boundary). 4). CD, DE along the brain expand or a straight line directly connected CE. 5). Different sizes for the extent of brain expansion (10, 15, or 20 mm). 6). Different sizes for the extent of skin flash (10, 15, or 20 mm). 7). The selection of the vertebral bodies C1 and C2. 8). The option of treatment includes the orbitals.}
  \label{fig:FieldOptions}
\end{figure}

The entire brain was considered to be the treatment target. As the eyes and lenses are sensitive to radiation, we needed to locate them and adjust the MLC to provide proper protection. Our proposed design is customizable and allows the exact relationship between the field aperture and these structures to be adjusted on the basis of local clinical requirements. The customizable options include the anterior-caudal boundary shapes, the extent of brain expansion, and the skin flash extension. In addition to these stylistic differences, the field shapes must also be adjustable based on the specifics of the patient’s disease. That is, patient-specific options, including the choice of having vertebral body C1 or C2 in the aperture and whether to include the orbitals. Different options for the visualization of the field shapes are shown in Figure ~\ref{fig:FieldOptions}. 

\subsection{Initial Field Aperture Selection}
\begin{figure}
  \centering
  \includegraphics[width=\textwidth]{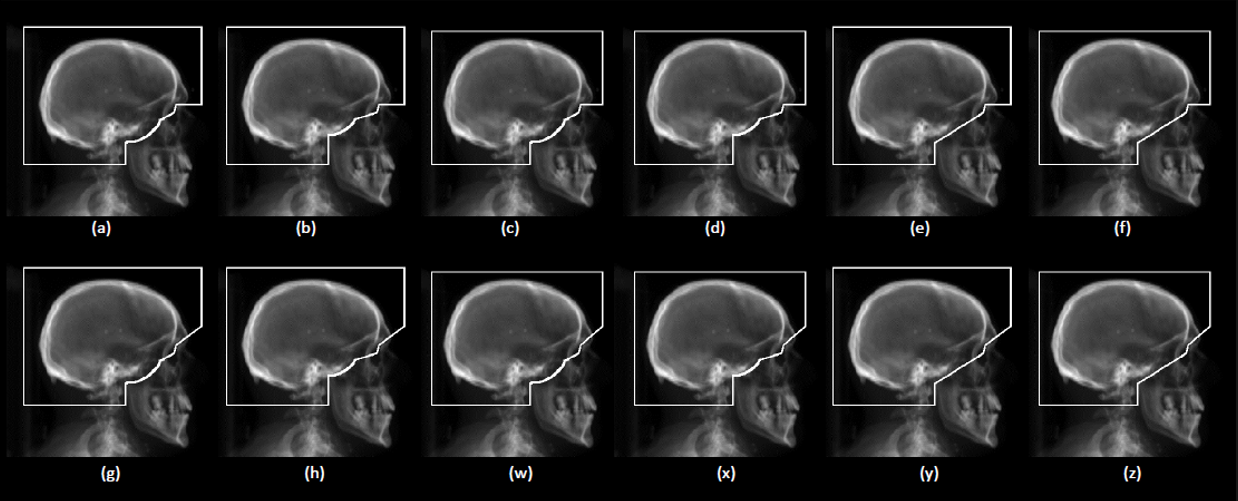}
  \caption{The initial field shape design was chosen from 12 candidate field shapes.}
  \label{fig:FieldReviewInitial}
\end{figure}

The initial field aperture was determined by seven radiation oncologists with a minimum of six years of experience and from five institutions worldwide. Evaluations were conducted using the scans of five patients. For each patient, 12 candidate field apertures (Figure ~\ref{fig:FieldReviewInitial}) were generated based on different combinations of the landmark-based options (options 1-6) shown in Figure ~\ref{fig:FieldOptions}. The reviewers ranked the top three field aperture shapes they preferred to use for each patient. A general score based on the five-point Likert scale (Table ~\ref{tab:LikertScale}) was given to each of the 12 candidate field shapes for each patient. The initial field aperture was then defined as the shape that received the highest rating and the highest scores. During the implementation of the field apertures based on the initial setting, clinicians’ feedback was collected each time an adjustment was made. 

\begin{table}[]
\caption{Five-point Likert scale for clinical reviews of field aperture design and plans.}
\begin{tabular}{p{0.02\textwidth}p{0.08\textwidth}p{0.8\textwidth}}%{lll}
\hline
\multicolumn{2}{l}{\textbf{Score}} & \textbf{Meaning   of Score in This Study}                                                                                                                                                                            \\ \hline
5    & Strongly Agree              & Use-as-is.   (i.e., clinically acceptable, and could be used for treatment without change)                                                                                                                           \\
4    & Agree                       & Minor   edits that are not necessary.    Stylistic but not clinically important differences. The current   contours/plan are acceptable.                                                                             \\
3    & Neither Agree or Disagree   & Minor   edits that are necessary.   Minor edits   are those that the reviewer judges can be made in less time than starting   from scratch or those that are expected to have minimal effect on treatment   outcome. \\
2    & Disagree                    & Major   edits. The necessary edits are required to ensure appropriate treatment, and   sufficiently significant that the user would prefer to start from scratch.                                                    \\
1    & Strongly Disagree           & Unusable.   The quality of the automatically generated contours or plan are so bad that   they are unusable.                                                                                                         \\ \hline
\end{tabular}\label{tab:LikertScale}

\end{table}

\subsection{Performance Evaluations}
We evaluated the performance of the field aperture designs and the generated plans using the initial field aperture settings. Scans from 182 patients from the clinic were included in the assessment of the quality of the generated field apertures. Thirty patients were randomly selected to create the treatment plans for further assessment, including dose coverage calculation and physician review.

\subsubsection{Quantitative Evaluation}
The quantitative evaluation metrics for field aperture design included Hausdorff Distance (HD) and Mean Surface Distance (MSD) which were computed to the clinical plans. Radiation doses to the structures were evaluated by calculating the maximum dose to the lenses, the mean dose to the eyes, and the brain volume receiving 95\% of the prescribed dose (V95\%). 

\subsubsection{Qualitative Evaluation}
To qualitatively assess the performance of the two approaches, physician reviews were performed on the design of the field apertures and the treatment plans. Five experienced radiation oncologists from four hospitals reviewed both the first and the second approaches using the plans generated from 30 patients. A five-point Likert scale (Table ~\ref{tab:LikertScale}) was used to determine clinical acceptability (scores $\geq$ 3).  Reviewers were asked to give a score for the field apertures and a score for the plans regarding the dose metrics including the isodose distribution on the transverse plane, the brain dose coverages, the maximum dose to the lens, the mean dose to the eyes, and the dose-volume histogram (DVH).   

\section{Results}
Clinicians’ feedback was collected each time an adjustment was made during the implementation of the field apertures. Then, the initial configuration (Figure ~\ref{fig:FieldOptions} (1)) was selected based on the field apertures that received the highest scores from the reviewers. The initial configuration included 1) the slash/diagonal-like AB line shape, 2) the CD, DE line shape along with the brain expansion, 3) the brain expansion sets 1.5 cm, 4) the skin flash sets 1.5cm, and 5) line BC at moderate distances between the backs of the lenses and eyes. 

\section{Quantitative Evaluation}

\begin{figure}
  \centering
  \includegraphics[width=\textwidth]{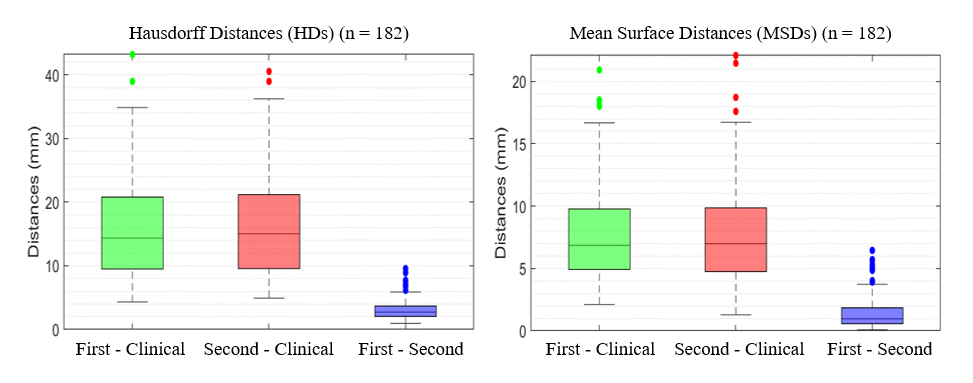}
  \caption{Box plots of the quantitative metrics of the field aperture design indicating the first and second approaches performed similarly.}
  \label{fig:FieldBoxPlot}
\end{figure}

\begin{figure}
  \centering
  \includegraphics[width=\textwidth]{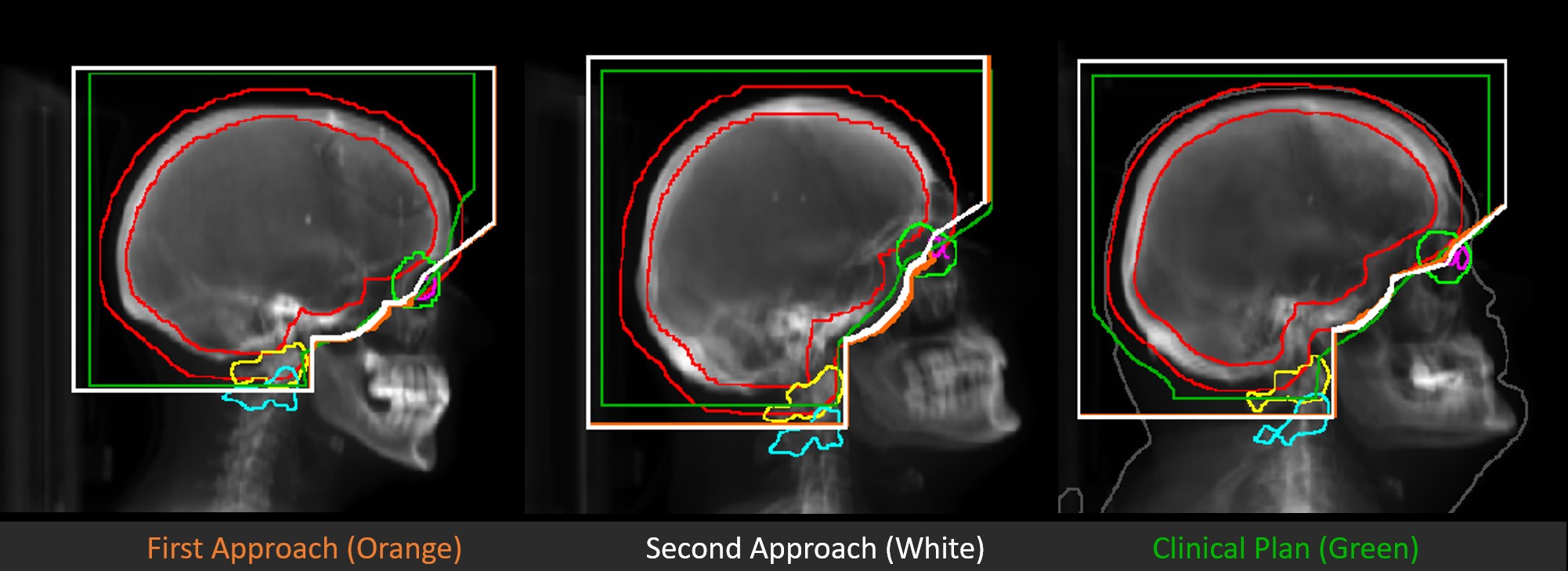}
  \caption{Comparisons of the resulting field apertures from the first (orange) and second (white) approaches and the clinical plans (green). Field apertures in orange and white are almost overlapped.}
  \label{fig:FieldFieldComparison}
\end{figure}

As shown in Figure ~\ref{fig:FieldBoxPlot}, the average HD and MSD for the generated field apertures were 16±7 mm and 7±3 mm, respectively, for the first approach, and 17±7 mm and 7±3 mm, respectively, for the second approach. The differences in HD and MSD between the first and the second approaches were 1±2 mm and 1±3 mm, respectively. Although the distances between the manual and automated approaches were fairly large, the majority of distances were contributed in less critical regions (Figure ~\ref{fig:FieldFieldComparison}). 

The different dose metrics compared between the clinical plans and the plans generated using the first and second approaches including the brain V95\%, the maximum doses to the lenses, and the mean dose to the eyes are shown in Table ~\ref{tab:DoseMetrics}. Both landmark-based approaches achieved a 100\% acceptance rate from the radiation oncologists for brain V95\%. The average acceptance rates for meeting lens dosimetric recommendations were 80\% (left lens) and 77\% (right lens) for the first approach, and 70\% (both left and right lenses) for the second approach, compared with 50\% (left lens) and 53\% (right lens) for the clinical plans, indicating that the automated plans maintained dose coverage of the brain with some reduction in the lens dose. 

% Please add the following required packages to your document preamble:
% \usepackage{multirow}
\begin{table}[]
\caption{Comparison of different dose metrics.}
\begin{tabular}{p{0.1\textwidth}p{0.09\textwidth}p{0.09\textwidth}p{0.06\textwidth}p{0.15\textwidth}p{0.15\textwidth}p{0.08\textwidth}p{0.08\textwidth}}%{llllllll}
\hline
\textbf{Target / OAR}       & \textbf{Plan Type} & \textbf{Dose / Volume Metric} & \textbf{Clinical Constraint} & \textbf{Resultant Dose} & \textbf{Clinical Dose}            & \textbf{Resultant Acceptance Rate (n=30)} & \textbf{Clinical Acceptance Rate (n=30)} \\ \hline
\multirow{2}{*}{Brain}      & First Approach     & V95\%                         & \textgreater{}95\%           & 99.97 ± 0.16\%          & \multirow{2}{*}{99.97 ± 0.17\%}   & 100\%                                     & \multirow{2}{*}{100\%}                   \\
                            & Second Approach    & V95\%                         & \textgreater{}95\%           & 99.97 ± 0.16\%          &                                   & 100\%                                     &                                          \\
\multirow{2}{*}{Left Lens}  & First Approach     & Max dose                      & \textless{}8Gy               & 5.90 ± 2.95 Gy          & \multirow{2}{*}{11.11 ± 9.24 Gy}  & 80\%                                      & \multirow{2}{*}{50\%}                    \\
                            & Second Approach    & Max dose                      & \textless{}8Gy               & 6.22 ± 3.62 Gy          &                                   & 70\%                                      &                                          \\
\multirow{2}{*}{Right Lens} & First Approach     & Max dose                      & \textless{}8Gy               & 6.57 ± 3.94 Gy          & \multirow{2}{*}{12.23 ± 10.11 Gy} & 77\%                                      & \multirow{2}{*}{53\%}                    \\
                            & Second Approach    & Max dose                      & \textless{}8Gy               & 6.65 ± 4.34 Gy          &                                   & 70\%                                      &                                          \\
\multirow{2}{*}{Left Eye}   & First Approach     & Mean dose                     & n.a.                         & 10.34 ± 1.85 Gy         & \multirow{2}{*}{13.21 ± 7.01 Gy}  & n.a.                                      & \multirow{2}{*}{n.a.}                    \\
                            & Second Approach    & Mean dose                     & n.a.                         & 9.85 ± 3.01 Gy          &                                   & n.a.                                      &                                          \\
\multirow{2}{*}{Right Eye}  & First Approach     & Mean dose                     & n.a.                         & 10.91 ± 1.93 Gy         & \multirow{2}{*}{13.71 ± 7.03 Gy}  & n.a.                                      & \multirow{2}{*}{n.a.}                    \\
                            & Second Approach    & Mean dose                     & n.a.                         & 10.38 ± 2.87 Gy         &                                   & n.a.                                      &                                          \\ \hline
\end{tabular}\label{tab:DoseMetrics}
\end{table}

\subsection{Qualitative Evaluation}
\begin{figure}
  \centering
  \includegraphics[width=\textwidth]{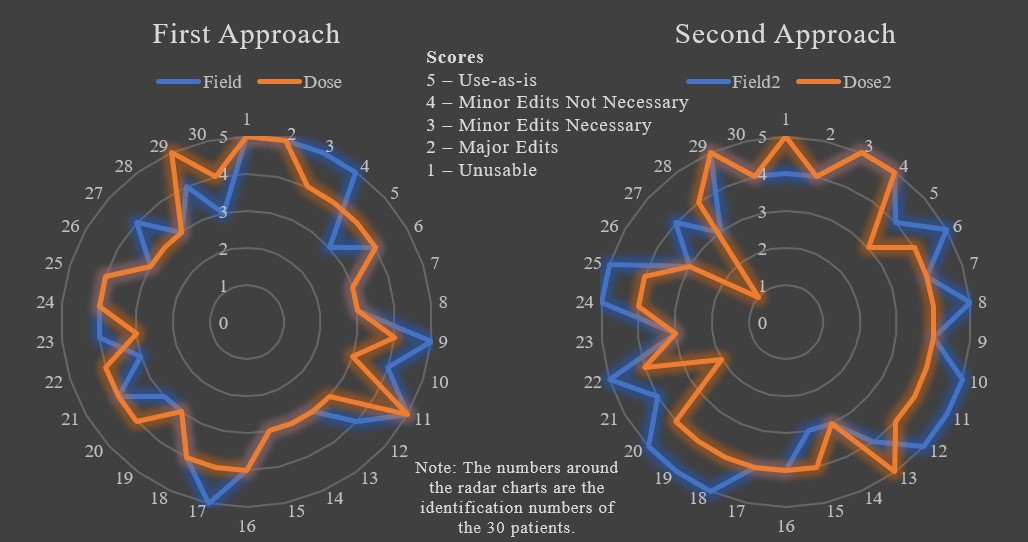}
  \caption{Radar plots of the clinical review results based on a five-point Likert scale.}
  \label{fig:ReviewResult}
\end{figure}

As shown in Figure ~\ref{fig:ReviewResult}, clinical reviews for the field aperture design and the plans were conducted on 30 patients using the five-point Likert scale. The field aperture designs created using the first and second approaches both achieved a 100\% acceptance rate, and the plans created using the first and second approaches achieved acceptance rates of 100\% and 93\%, respectively. 

The field apertures generated using the first and second approaches were compared with each other and with the field apertures in the clinical plans. As shown in Figure ~\ref{fig:FieldFieldComparison}, the field apertures generated using the first and second approaches were similar and agreed reasonably well with those in the clinical plans, especially for the regions from the eyes to the vertebral bodies. 

\section{Discussion}
In this work, we have presented a novel pipeline for automatically generating WBRT treatment field apertures using anatomical landmarks. The pipeline consisted of two approaches to obtaining the anatomical landmarks: the first approach was based on 3D CT segmentation, and the second approach was based on 2D DRR segmentation. We successfully generated field apertures from both approaches, and the generated field apertures were similar to those in the clinical plans and achieved high clinical acceptance rates. 

The design of the two parallel field aperture generation approaches could be an effective tool for quality control, as the resulting field apertures were very similar. In practice, if the first approach fails to generate the field apertures, the second approach could be used to calculate the distances of the field apertures. An HD distance of 14 mm should be used to indicate a pass or fail in the automatic beam aperture check. Our future studies will further investigate the criteria for and the effectiveness of identifying automated field aperture check failures following the quality assurance approaches of Rhee et al. and Kisling et al.~\cite{rhee2019automatic,kisling2020automatic}

\begin{figure}
  \centering
  \includegraphics[width=0.75\textwidth]{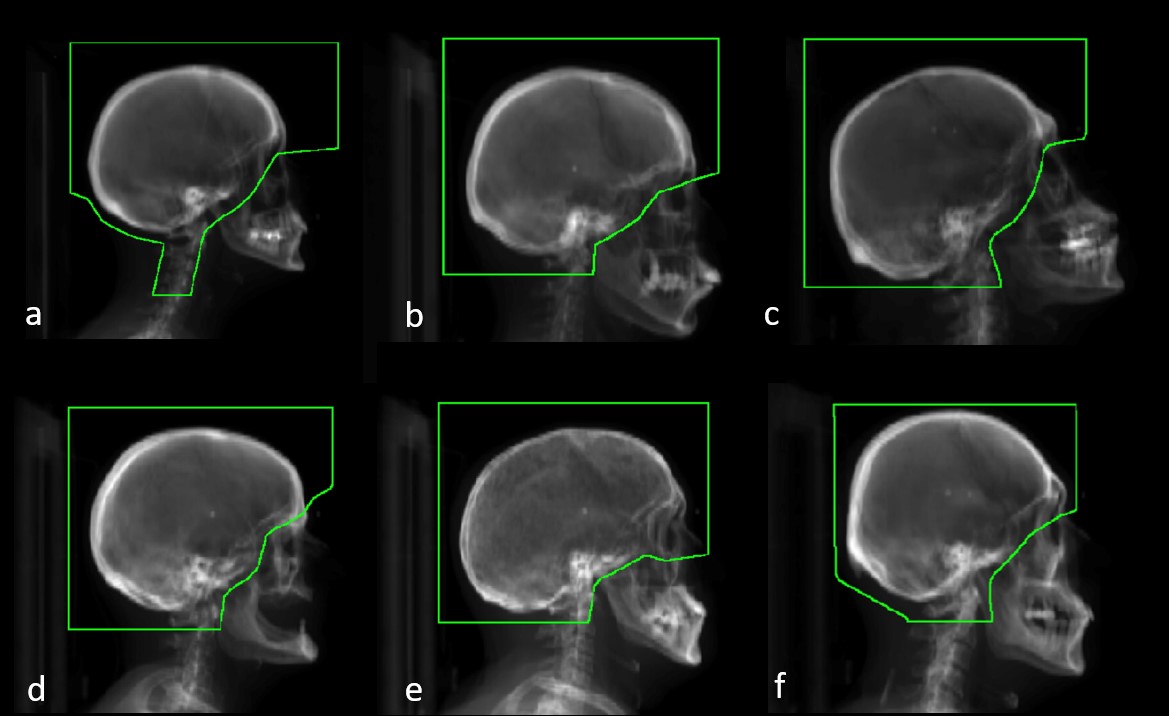}
  \caption{Clinical variations in the field aperture design.}
  \label{fig:FieldVariations}
\end{figure}

Field aperture designs vary according to the different purposes for which they are used and preferences in different clinical practices (Figure ~\ref{fig:FieldVariations}). Our landmark-based field aperture generation method can be easily configured to suit different clinical purposes and clinical styles. In Figure ~\ref{fig:FieldDesign}, we detailed how we selected the landmarks to form the boundaries of the field apertures. By adjusting the positions and the shapes of these lines, we allow the flexibility in defining different field aperture shapes to address the needs of different clinical practices. For example, in our hospital, clinicians preferentially use field aperture shapes described as the initial setup (Figure ~\ref{fig:FieldOptions}, the initial setting), while other hospitals may prefer to use a different setting (Figure ~\ref{fig:FieldOptions} option 2-6). Our landmark-based method provides a feasible way of changing the field aperture configurations by introducing user-specific options.

The automated WBRT treatment planning pipeline ensures the radiation plan generation is fully automated without human interaction during the planning process. The automation is beneficial for both clinicians and patients, where we can reduce clinicians’ workloads and shorten the treatment planning time down to a few minutes. We have implemented this automated landmark-based WBRT treatment planning as part of automated radiotherapy planning tools. In the future, we plan to develop a method to predict and automatically adjust specific field shapes by learning from the experience and preferred shapes in clinicians’ routine practices.

\section{Conclusion}
In this work, we developed and evaluated a novel pipeline consisting of two landmark-based field aperture generation approaches for WBRT treatment planning; they are fully automated and customizable. The automation pipeline is beneficial for both clinicians and patients, where we can reduce clinicians’ workload and reduce the treatment planning time. The customizability of the field aperture design addresses different clinical requirements and allows the personalized design to become feasible. The performance results regarding quantitative and qualitative evaluations demonstrated that our plans were comparable with the original clinical plans. This technique has been deployed as part of a fully automated treatment planning tool for whole-brain cancer and could be translated to other treatment sites in the future. 

\section*{Author Contributions}
Yao Xiao – writing the manuscript, second approach ROI segmentation deep learning model development, customizable landmark-based field aperture design and development, data analysis; Carlos Cardenas – data collection, customizable landmark-based field aperture design, ROI projection code development, supervising research, manuscript review; Dong Joo Rhee – first approach ROI segmentation deep learning model development, manuscript review; Tucker Netherton – first approach ROI segmentation deep learning model development, manuscript review; Lifei Zhang – DRR generation code development, RPA deployment, manuscript review; Callistus Nguyen – deep learning models deployment, clinical integration, manuscript review; Raphael Douglas – dose data collection, RPA deployment, manuscript review; Raymond Mumme – data collection, manuscript review; Stephen Skett – clinical integration, manuscript review; Tina Patel – clinical integration, manuscript review; Chris Trauernicht – clinical integration, manuscript review; Caroline Chung – clinical review, clinical integration, manuscript review; Hannah Simonds – clinical review, clinical integration, manuscript review; Ajay Aggarwal – clinical review, clinical integration, manuscript review; Laurence Court – writing manuscript, supervising research, manuscript review. 

\section*{Acknowledgments}
The authors highly appreciate Caroline Chung, Prajnan Das, Anuja Jhingran, Simeng Zhu, Melody Xu, Indranil Mallick, Mary Feng, and Hina Saeed for reviewing the field aperture design and the generated plans and providing prestigious feedback during the implementation of the field apertures. The authors thank Laura Russell from the department of scientific publications at The University of Texas MD Anderson Cancer Center for editing this work. The authors would also like to thank the Wellcome Trust for funding this work. We gratefully acknowledge the support of The University of Texas MD Anderson Cancer Center High-Performance Computing Center for computational resources to support the present research.

%Bibliography
\bibliographystyle{unsrt}  
\bibliography{LandmarkWBRT}

\begin{thebibliography}{10}

\bibitem{sung2021global}
Hyuna Sung, Jacques Ferlay, Rebecca~L Siegel, Mathieu Laversanne, Isabelle
  Soerjomataram, Ahmedin Jemal, and Freddie Bray.
\newblock Global cancer statistics 2020: Globocan estimates of incidence and
  mortality worldwide for 36 cancers in 185 countries.
\newblock {\em CA: a cancer journal for clinicians}, 71(3):209--249, 2021.

\bibitem{sperduto2020survival}
Paul~W Sperduto, Shane Mesko, Jing Li, Daniel Cagney, Ayal Aizer, Nancy~U Lin,
  Eric Nesbit, Tim~J Kruser, Jason Chan, Steve Braunstein, et~al.
\newblock Survival in patients with brain metastases: summary report on the
  updated diagnosis-specific graded prognostic assessment and definition of the
  eligibility quotient.
\newblock {\em Journal of Clinical Oncology}, 38(32):3773--3784, 2020.

\bibitem{tsao2012radiotherapeutic}
May~N Tsao, Dirk Rades, Andrew Wirth, Simon~S Lo, Brita~L Danielson, Laurie~E
  Gaspar, Paul~W Sperduto, Michael~A Vogelbaum, Jeffrey~D Radawski, Jian~Z
  Wang, et~al.
\newblock Radiotherapeutic and surgical management for newly diagnosed brain
  metastasis (es): An american society for radiation oncology evidence-based
  guideline.
\newblock {\em Practical radiation oncology}, 2(3):210--225, 2012.

\bibitem{agazaryan2020timeliness}
Nzhde Agazaryan, Phillip Chow, James Lamb, Minsong Cao, Ann Raldow, Phillip
  Beron, John Hegde, and Michael Steinberg.
\newblock The timeliness initiative: continuous process improvement for prompt
  initiation of radiation therapy treatment.
\newblock {\em Advances in Radiation Oncology}, 5(5):1014--1021, 2020.

\bibitem{shah2019cancer}
Shailja~C Shah, Violet Kayamba, Richard~M Peek~Jr, and Douglas Heimburger.
\newblock Cancer control in low-and middle-income countries: is it time to
  consider screening?
\newblock {\em Journal of global oncology}, 5:1--8, 2019.

\bibitem{kao2015tumor}
Johnny Kao, Boramir Darakchiev, Linda Conboy, Sara Ogurek, Neha Sharma, Xuemin
  Ren, and Jeffrey Pettit.
\newblock Tumor directed, scalp sparing intensity modulated whole brain
  radiotherapy for brain metastases.
\newblock {\em Technology in cancer research \& treatment}, 14(5):547--555,
  2015.

\bibitem{jiang2019dosimetric}
Aijun Jiang, Weipeng Sun, Fen Zhao, Zhenxuan Wu, Dongping Shang, Qingxi Yu,
  Suzhen Wang, Jian Zhu, Fengchang Yang, and Shuanghu Yuan.
\newblock Dosimetric evaluation of four whole brain radiation therapy
  approaches with hippocampus and inner ear avoidance and simultaneous
  integrated boost for limited brain metastases.
\newblock {\em Radiation Oncology}, 14(1):1--8, 2019.

\bibitem{lin2020automated}
Ting-Chun Lin, Chih-Yuan Lin, Kai-Chiun Li, Jin-Huei Ji, Ji-An Liang, An-Cheng
  Shiau, Liang-Chih Liu, and Ti-Hao Wang.
\newblock Automated hypofractionated imrt treatment planning for early-stage
  breast cancer.
\newblock {\em Radiation Oncology}, 15(1):1--9, 2020.

\bibitem{han2021clinical}
Eun~Young Han, Carlos~E Cardenas, Callistus Nguyen, Donald Hancock, Yao Xiao,
  Raymond Mumme, Laurence~E Court, Dong~Joo Rhee, Tucker~J Netherton, Jing Li,
  et~al.
\newblock Clinical implementation of automated treatment planning for
  whole-brain radiotherapy.
\newblock {\em Journal of applied clinical medical physics}, 22(9):94--102,
  2021.

\bibitem{rhee2019automatic}
Dong~Joo Rhee, Carlos~E Cardenas, Hesham Elhalawani, Rachel McCarroll, Lifei
  Zhang, Jinzhong Yang, Adam~S Garden, Christine~B Peterson, Beth~M Beadle, and
  Laurence~E Court.
\newblock Automatic detection of contouring errors using convolutional neural
  networks.
\newblock {\em Medical physics}, 46(11):5086--5097, 2019.

\bibitem{kisling2020automatic}
Kelly Kisling, Carlos Cardenas, Brian~M Anderson, Lifei Zhang, Anuja Jhingran,
  Hannah Simonds, Peter Balter, Rebecca~M Howell, Kathleen Schmeler, Beth~M
  Beadle, et~al.
\newblock Automatic verification of beam apertures for cervical cancer
  radiation therapy.
\newblock {\em Practical radiation oncology}, 10(5):e415--e424, 2020.

\bibitem{netherton2020evaluation}
Tucker~J Netherton, Dong~Joo Rhee, Carlos~E Cardenas, Caroline Chung, Ann~H
  Klopp, Christine~B Peterson, Rebecca~M Howell, Peter~A Balter, and Laurence~E
  Court.
\newblock Evaluation of a multiview architecture for automatic vertebral
  labeling of palliative radiotherapy simulation ct images.
\newblock {\em Medical physics}, 47(11):5592--5608, 2020.

\bibitem{long2015fully}
Jonathan Long, Evan Shelhamer, and Trevor Darrell.
\newblock Fully convolutional networks for semantic segmentation.
\newblock In {\em Proceedings of the IEEE conference on computer vision and
  pattern recognition}, pages 3431--3440, 2015.

\end{thebibliography}

\end{document}